# Bianchi type-I dark energy cosmology with power-law relation in Brans-Dicke theory of gravitation

S.D. Katore[1], D.V. Kapse[2]

[1] Department of Mathematics, SGBAU, Amravati-444 602, India.

E-mail: katoresd@rediffmail.com

[2] Department of Mathematics, PRMIT&R, Badnera-Amravati-444 701, India.

E-mail: dipti.kapse@gmail.com

**Abstract:** In this paper, we have studied the interacting and non-interacting dark energy and dark matter in the spatially homogenous and anisotropic Bianchi type-I model in the Brans-Dicke theory of gravitation. The field equations have been solved by using (i) power-law relation and (ii) by assuming scale factor in terms of redshift. Here we have considered two cases of an interacting and non-interacting dark energy scenario and obtained general results. It has been found that for suitable choice of interaction between dark energy and dark matter we can avoid the coincidence problem which appears in the $\Lambda CDM$ model. Some physical aspects and stability of the models are discussed in detail. The statefinder diagnostic pair i.e. $\{r, s\}$ is adopted to differentiate our dark energy models.

## 1. Introduction

The recent cosmological observational data of Type Ia Supernovae (SNeIa) (Riess et al. [1]; Perlmutter et al. [2]), Cosmic Microwave Background (CMB) (Bennett et al. [3]; Spergel et al.[4]), Large Scale Structure (LSS) (Tegmark et al. [5,6]), the Sloan Digital Sky Survey (SDSS) (Seljak et al.[7], Adeleman-McCarthy et al. [8]), Wilkinson Microwave Anisotropy Probe (WMAP) (Knop et al. [9]) and Chandra X-ray observatory (Allen et al. [10]), it strongly suggests that our universe is dominated by a component with large negative pressure called as dark energy (DE).

The study of DE is possible through its equation of state (EoS) parameter $\omega^{de} = p^{de}/\rho^{de}$ which is not necessarily constant, where $p^{de}$ is the pressure and $\rho^{de}$ is the energy density of DE. The DE candidate which can simply explain the cosmic acceleration is a vacuum energy $\left(\omega^{de}=-1\right)$, which is mathematically equivalent to the cosmological constant ($\Lambda$). The other conventional alternatives, which can be described by minimally coupled scalar fields, are quintessence $\left(-1<\omega^{de}<-\frac{1}{3}\right)$, phantom $\left(\omega^{de}<-1\right)$ and quintom (that can across from phantom region to quintessence region). From observational results coming from SNe Ia data (Knop et al. [9]) and combination of SNe Ia data with CMBR anisotropy and galaxy clustering statistics (Tegmark et al. [8] ), the limits on EoS parameter are obtained as $-1.67<\omega^{de}<-0.62$ and $-1.33<\omega^{de}<-0.79$ respectively. Recently, DE models with variable EoS parameter have been studied by Ram et al. [11, 12], Katore et al. [13], Reddy et al. [14] and Mahanta et al. [15].

Interaction between DE and DM lead to a solution to the coincidence problem (Cimento et al. [16]; Dalal et al. [17]; Jamil and Rashid [18, 19]). By considering a coupling between DE and DM, we can explain why the energy densities of DE and DM are nearly equal today. Due to interaction between two components, the energy conservation can't hold for the individual components. Recent observations (Bertolami et al. [20]; Le Delliou et al. [21]; Berger and Shojaei [22]) provide the evidence for the possibility of such an interaction between DE and DM. Zhang [23, 24], Zimdahl and Pavon [25], Pradhan et al. [26, 27], Saha et al. [28], Amirhashchi et al.[29-33], Adhav et al.[34, 35], Fayaz [36] have investigated various cosmological models with interacting DE.

The Brans-Dicke theory [37] is a generalized form of general relativity and it is one of the most enchanting examples of scalar tensor theories of gravitation. Brans-Dicke (BD) theory introduces a scalar field $\phi$ which has the dimensions of the inverse of gravitational constant and which interacts equally with all forms of matter. Recently, Rao et al. [38], Sarkar [39, 40], Katore et al. [41], Singh and Dewri [42] and Reddy et al. [43, 44] have studied the cosmological models in Brans-Dicke theory of gravitation.

Amirhashchi et al.[33] have investigated the DE equation of state (EoS) parameter in both interacting and non-interacting cases and examined its future by applying hyperbolic scale factor in general relativity. Motivated by the above investigations, in this paper, we have extended the work of Amirhashchi et al. [33] in Brans-Dicke theory of gravitation. This is relevant because of the fact that scalar field plays an important role in the discussion of DE models. In this paper, we have studied the interacting and non-interacting dark energy and dark matter in the spatially homogenous and anisotropic Bianchi type-I model in the Brans-Dicke theory of gravitation. The field equations have been solved by using (i) power-law relation and (ii) by assuming scale factor in terms of redshift and have discussed the physical properties and also the physical acceptability and stability of our models. The values of cosmological parameters are taken from the recent observations made by Amirhashchi [45], Amirhashchi and Amirhashchi [46, 47] and Patrignani et al. [48]. The paper has following structure. In section 2, the metric and the Brans-Dicke field equations are described. Section 3 is devoted to the solution of the field equations. Using the scale factor as a function of redshift, we have obtained our results for non-interacting and interacting cases. In Section 4, we have discussed the physical aspects and stability of models. In section 5, behavior of anisotropy parameter of expansion $(\Delta)$ is studied. The statefinder diagnostic pair i.e. {$r$, $s$} is adopted to characterize different phases of the universe in section 6 and finally, Section 7 contains some concluding remarks.

## 2. The metric and BD field equations

We consider the homogeneous and anisotropic Bianchi type-I universe as

$$ds^2 = -dt^2 + A^2 dx^2 + B^2 dy^2 + C^2 dz^2, \qquad (1)$$

where the scale factors $A, B$ and $C$ are functions of time $t$ only.

BD field equations for the combined scalar and tensor fields with $(8\pi G = c = 1)$ are given by (Brans-Dicke [37], Reddy et al.[43], Rao et al.[49])

$$R_{ij} - \frac{1}{2} g_{ij} R - \bar{\omega} \phi^{-2} \left( \phi_{,i} \phi_{,j} - \frac{1}{2} g_{ij} \phi^{,k} \phi_{,k} \right) - \phi^{-1} \left( \phi_{i;j} - g_{ij} \phi^{,k}_{;k} \right) = \phi^{-1} (T_{ij}), \qquad (2)$$

where $R$ is the Ricci scalar, $R_{ij}$ is the Ricci tensor, $\phi$ is the Brans-Dicke scalar field, $\bar{\omega}$ is the dimensionless constant and $T_{ij}$ is the energy momentum tensor. The scalar fields satisfy the following equation

$$\phi^{,k}_{;k} = \frac{T^i_j}{3 + 2\bar{\omega}}. \qquad (3)$$

The energy momentum tensor is given by

$$T^i_j = T^{(m)i}_j + T^{(de)i}_j, \qquad (4)$$

where $T_j^{(m)i}$ and $T_j^{(de)i}$ are energy momentum tensors of dark matter and dark energy, respectively. These are given by

$$T_j^{(m)i} = diag\left[-\rho^m, p^m, p^m, p^m\right],$$
$$= diag\left[-1, \omega^m, \omega^m, \omega^m\right]\rho^m \tag{5}$$

and

$$T_j^{(de)i} = diag\left[-\rho^{de}, p^{de}, p^{de}, p^{de}\right],$$
$$= diag\left[-1, \omega^{de}, \omega^{de}, \omega^{de}\right]\rho^{de}, \tag{6}$$

where $\rho^m$ and $\rho^{de}$ are energy densities of DM and DE respectively. Similarly, $p^m$ and $p^{de}$ are the pressure of DM and DE respectively, while $\omega^m = p^m/\rho^m$ and $\omega^{de} = p^{de}/\rho^{de}$ are the corresponding EoS parameters of DM and DE (Harko et al. [50]).

In co-moving coordinate system, the BD field equations (2) and (3) for the metric (1), are given by

$$\frac{\dot{A}\dot{B}}{AB} + \frac{\dot{B}\dot{C}}{BC} + \frac{\dot{A}\dot{C}}{AC} - \frac{\overline{\omega}}{2}\frac{\dot{\phi}^2}{\phi^2} + \left(\frac{\dot{A}}{A} + \frac{\dot{B}}{B} + \frac{\dot{C}}{C}\right)\frac{\dot{\phi}}{\phi} = \frac{\rho^m + \rho^{de}}{\phi}, \tag{7}$$

$$\frac{\ddot{B}}{B} + \frac{\ddot{C}}{C} + \frac{\dot{B}\dot{C}}{BC} + \frac{\overline{\omega}}{2}\frac{\dot{\phi}^2}{\phi^2} + \frac{\ddot{\phi}}{\phi} + \left(\frac{\dot{B}}{B} + \frac{\dot{C}}{C}\right)\frac{\dot{\phi}}{\phi} = \frac{1}{\phi}\left[-\omega^m \rho^m - \omega^{de}\rho^{de}\right], \tag{8}$$

$$\frac{\ddot{A}}{A} + \frac{\ddot{C}}{C} + \frac{\dot{A}\dot{C}}{AC} + \frac{\overline{\omega}}{2}\frac{\dot{\phi}^2}{\phi^2} + \frac{\ddot{\phi}}{\phi} + \left(\frac{\dot{A}}{A} + \frac{\dot{C}}{C}\right)\frac{\dot{\phi}}{\phi} = \frac{1}{\phi}\left[-\omega^m \rho^m - \omega^{de}\rho^{de}\right], \tag{9}$$

$$\frac{\ddot{A}}{A} + \frac{\ddot{B}}{B} + \frac{\dot{A}\dot{B}}{AB} + \frac{\overline{\omega}}{2}\frac{\dot{\phi}^2}{\phi^2} + \frac{\ddot{\phi}}{\phi} + \left(\frac{\dot{A}}{A} + \frac{\dot{B}}{B}\right)\frac{\dot{\phi}}{\phi} = \frac{1}{\phi}\left[-\omega^m \rho^m - \omega^{de}\rho^{de}\right], \tag{10}$$

and the wave equation is

$$\ddot{\phi} + \left(\frac{\dot{A}}{A} + \frac{\dot{B}}{B} + \frac{\dot{C}}{C}\right)\dot{\phi} = \frac{\rho^m(1-3\omega^m) + \rho^{de}(1-3\omega^{de})}{3+2\overline{\omega}}, \tag{11}$$

where an overhead dot denotes differentiation with respect to *t*.

### 3. Solutions of field equations

We have initially six variables and four linearly independent equations (7)-(10). The system is thus initially undetermined and we need additional condition to solve the system completely. In order to solve these field equations, we first assume the power-law relation between the average scale factor ($a$) and scalar field ($\phi$) (Pimental [51], Johri and Desikan [52]) as

$$\phi = \alpha\, a^{\beta}, \tag{12}$$

where $\alpha$ and $\beta > 0$ are constants.

To examine the general results, we assume that the average scale factor is a hyperbolic function of time as Amirhashchi [53]

$$a(t) = \sinh(t),$$

which gives dynamical deceleration parameter ($q$). Also, Chen and Kao [54] have shown that this scale factor is stable under metric perturbation.

In terms of redshift the above scale factor is given by

$$a(t) = \frac{1}{1+z}, \quad z = \frac{1}{\sinh(t)} - 1, \tag{13}$$

where $z$ is the redshift parameter and $a = (ABC)^{1/3}$ is the average scale factor.

From equations (8)-(10), we obtain

$$\frac{\ddot{A}}{A} - \frac{\ddot{B}}{B} + \frac{\dot{C}}{C}\left(\frac{\dot{A}}{A} - \frac{\dot{B}}{B}\right) + \left(\frac{\dot{A}}{A} - \frac{\dot{B}}{B}\right)\frac{\dot{\phi}}{\phi} = 0, \tag{14}$$

$$\frac{\ddot{B}}{B} - \frac{\ddot{C}}{C} + \frac{\dot{A}}{A}\left(\frac{\dot{B}}{B} - \frac{\dot{C}}{C}\right) + \left(\frac{\dot{B}}{B} - \frac{\dot{C}}{C}\right)\frac{\dot{\phi}}{\phi} = 0, \tag{15}$$

$$\frac{\ddot{A}}{A} - \frac{\ddot{C}}{C} + \frac{\dot{B}}{B}\left(\frac{\dot{A}}{A} - \frac{\dot{C}}{C}\right) + \left(\frac{\dot{A}}{A} - \frac{\dot{C}}{C}\right)\frac{\dot{\phi}}{\phi} = 0. \tag{16}$$

Solving equations (14)-(16), we obtain

$$\frac{\dot{A}}{A} - \frac{\dot{B}}{B} = \frac{k_1}{ABC\phi}, \tag{17}$$

$$\frac{\dot{B}}{B} - \frac{\dot{C}}{C} = \frac{k_2}{ABC\phi}, \tag{18}$$

$$\frac{\dot{A}}{A} - \frac{\dot{C}}{C} = \frac{k_3}{ABC\phi}, \tag{19}$$

where $k_1, k_2$ and $k_3$ are constants of integration.

Equations (17)-(19) further reduces to

$$\frac{A}{B} = d_1 \exp\left(k_1 \int \frac{dt}{ABC\phi}\right), \tag{20}$$

$$\frac{B}{C} = d_2 \exp\left(k_2 \int \frac{dt}{ABC\phi}\right), \qquad (21)$$

$$\frac{A}{C} = d_3 \exp\left(k_3 \int \frac{dt}{ABC\phi}\right), \qquad (22)$$

where $d_1, d_2$ and $d_3$ are constants of integration.

Using equations (20)-(22), we can write the metric functions *A, B* and *C* explicitly as

$$A = a_1 a \exp\left(b_1 \int \frac{dt}{a^3 \phi}\right), \qquad (23)$$

$$B = a_2 a \exp\left(b_2 \int \frac{dt}{a^3 \phi}\right), \qquad (24)$$

$$C = a_3 a \exp\left(b_3 \int \frac{dt}{a^3 \phi}\right), \qquad (25)$$

where $a_1 = (d_1 d_2)^{\frac{1}{3}}, a_2 = (d_1^{-1} d_3)^{\frac{1}{3}}, a_3 = (d_2 d_3)^{-\frac{1}{3}}, b_1 = \frac{k_1 + k_2}{3}, b_2 = \frac{k_3 - k_1}{3}, b_3 = -\frac{k_2 + k_3}{3}$,

which satisfies the relations $a_1 a_2 a_3 = 1$ and $b_1 + b_2 + b_3 = 0$.

Using equations (23)-(25) in (7), we obtain

$$H^2 = \frac{\rho^m + \rho^{de}}{D\phi} + \frac{K}{D} a^{-6} \phi^{-2}, \qquad (26)$$

where $K = -(b_1 b_2 + b_2 b_3 + b_1 b_3)$, $D = \left(3 + \beta - \beta^2 \frac{\overline{\omega}}{2}\right)$ and $H = \frac{1}{3}\left(\frac{\dot{A}}{A} + \frac{\dot{B}}{B} + \frac{\dot{C}}{C}\right)$ is the Hubble parameter. For $b_1 = b_2 = b_3 = 0,$ the model reduces to the flat FRW model in BD theory.

The energy conservation equation $T_{;j}^{ij} = 0$ is $T^{(m)\,ij}_{\quad;j} + {}^{de}T^{(de)\,ij}_{\quad;j} = 0$ and is given by

$$\dot{\rho}^m + 3\frac{\dot{a}}{a}(1 + \omega^m)\rho^m + \dot{\rho}^{de} + 3\frac{\dot{a}}{a}(1 + \omega^{de})\rho^{de} = 0. \qquad (27)$$

### 3.1. Non-Interacting Dark energy and Dark matter

In this section we have considered that there is no interaction between DE and DM. Therefore, the general form of energy conservation equation (27), leads to (Harko et al. [50])

$$\dot{\rho}^m + 3\frac{\dot{a}}{a}(1 + \omega^m)\rho^m = 0 \qquad (28)$$

and

$$\dot{\rho}^{de} + 3\frac{\dot{a}}{a}(1 + \omega^{de})\rho^{de} = 0. \qquad (29)$$

Using equation (28), we obtain the energy density of DM as

$$\rho^m = \rho_0^m a^{-3(1+\omega^m)} \tag{30}$$

$$= \rho_0^m (1+z)^{3(1+\omega^m)}, \tag{31}$$

where $\rho_0^m > 0$ is a constant of integration.

Using equations (30) and (26), we obtain the energy density of DE as

$$\rho^{de} = DH^2 \alpha\, a^\beta - \rho_0^m a^{-3(1+\omega^m)} - K\alpha^{-1} a^{-(\beta+6)} \tag{32}$$

$$= \left(1+(1+z)^2\right)\left(D\alpha(1+z)^{-\beta} - 3\Omega_0^m (1+z)^{3(1+\omega^m)}\right) - K\alpha^{-1}(1+z)^{(\beta+6)},$$

where $\Omega^m = \dfrac{\rho^m}{3H^2}$ is the energy density of DM and 0 denotes the present value of $\Omega^m$.

Using equations (12), (30) and (32) in equation (8), we obtain the EoS parameter of DE as

$$\omega^{de} = \frac{H^2\left[(\beta+2)q - \left(\beta^2\left(1+\dfrac{\overline{\omega}}{2}\right)+\beta+1\right)\right]\alpha\, a^\beta + K\alpha^{-1} a^{-(\beta+6)} - 3\omega^m \Omega_0^m H^2 a^{-3(1+\omega^m)}}{H^2\left(D\alpha\, a^\beta - 3\Omega_0^m a^{-3(1+\omega^m)}\right) - K\alpha^{-1} a^{-(\beta+6)}}. \tag{33}$$

where $q = -\dfrac{\ddot{a}}{aH^2} = -\dfrac{1}{1+(1+z)^2}$ is the deceleration parameter.

Using equation (13) in equation (33), we obtain the EoS parameter of DE in terms of redshift as

$$\omega^{de} = -\frac{\left[(\beta+2)\dfrac{1}{1+(1+z)^2} + \left(\beta^2\left(1+\dfrac{\overline{\omega}}{2}\right)+\beta+1\right)\right]\alpha(1+z)^{-\beta} - K\alpha^{-1}\dfrac{(1+z)^{(\beta+6)}}{1+(1+z)^2} + 3\omega^m \Omega_0^m (1+z)^{3(1+\omega^m)}}{D\alpha(1+z)^{-\beta} - K\alpha^{-1}\dfrac{(1+z)^{(\beta+6)}}{1+(1+z)^2} - 3\Omega_0^m (1+z)^{3(1+\omega^m)}}.$$

(34)

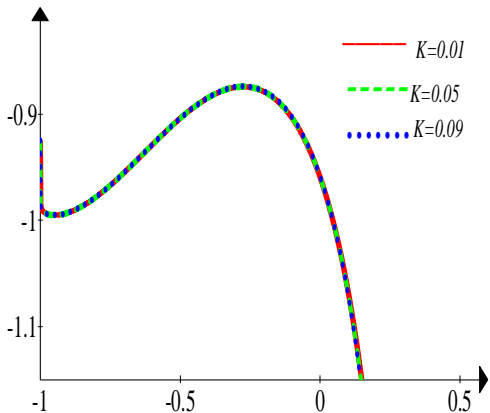

Fig.1: The plot of EoS parameter $\omega^{de}$ versus redshift $z$ for $\alpha=1, \beta=0.01, \Omega_0^m=0.3$, $\omega^m=0, \overline{\omega}=2$ and vary K=0.01, 0.05, 0.09

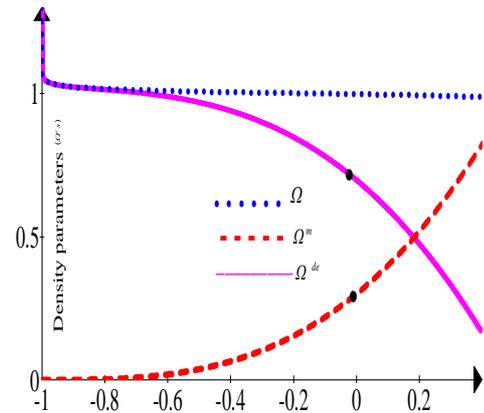

Fig.2: The plot of total energy densities versus redshift $z$ for $\alpha=1, \beta=0.01, \Omega_0^m=0.3$, $\omega^m=0, \overline{\omega}=2$ and K=0.01.

resp.

The behavior of EoS parameter of DE in terms of redshift $z$ is depicted in Fig. 1. Here the parameter $\omega^m$ is taken to be zero and vary constant $K$ as 0.01, 0.05 and 0.09 respectively. From figure it is clear that for all small values of $K$, the EoS of DE is varying in quintessence region and crossing Phantom Divide line (PDL) $\omega^{de} = -1$. However, it is observed that at late time (i.e. at $z = -1$) the EoS parameter $\omega^{de} \approx -1$. Therefore we say that the cosmological constant is a suitable candidate to represent the behavior of DE in the derived model at late times.

The matter energy density parameter $\Omega^m$ and dark energy density parameter $\Omega^{de}$ are given by

$$\Omega^m = \frac{\rho^m}{3H^2}$$
$$= \Omega_0^m (1+z)^{3(1+\omega^m)}, \tag{35}$$

$$\Omega^{de} = \frac{\rho^{de}}{3H^2}$$
$$= \frac{D\alpha}{3}(1+z)^{-\beta} - \frac{K\alpha^{-1}(1+z)^{(\beta+6)}}{3(1+(1+z)^2)} - \Omega_0^m(1+z)^{3(1+\omega^m)}. \tag{36}$$

Using equations (35) and (36), we obtain overall density parameter $\Omega$ as

$$\Omega = \Omega^m + \Omega^{de}$$
$$= \frac{D\alpha}{3}(1+z)^{-\beta} - \frac{K\alpha^{-1}(1+z)^{(\beta+6)}}{3(1+(1+z)^2)}. \tag{37}$$

The variation of density parameters $\Omega^m$ and $\Omega^{de}$ with redshift $z$ is depicted in Fig. 2. Dot denotes the current value of these parameters. It is observed that for sufficiently large time, the overall density parameter ($\Omega$) approaches to 1. Therefore the model predicts a flat universe at late time. It is interesting to note that the value of $\beta$ (i. e. BD theory) brings impact on the evolution of the densities (Fig. 2). The dark energy density parameter is increasing whereas the matter density parameter is decreasing. It is also clear that the value of dark energy density parameter is greater than the matter density parameter. Thus, the universe is dominated by dark energy throughout the evolution.

**3.2. Interacting Dark energy and Dark matter**

In this section, we have considered interaction between DE and DM. For this purpose we can write the energy conservation equation (27) as

$$\dot{\rho}^m + 3\frac{\dot{a}}{a}(1+\omega^m)\rho^m = Q \tag{38}$$

and

$$\dot{\rho}^{de} + 3\frac{\dot{a}}{a}(1+\omega^{de})\rho^{de} = -Q, \tag{39}$$

where $Q$ is interacting term.

To find the solution of coincidence problem, we have considered an energy transfer from dark energy to dark matter by assuming $Q>0$, which ensures that the second law of thermodynamics is fulfilled (Pavon and Wang [55]). The continuity equation (38) and (39) implies that the interaction term $Q$ should be proportional to inverse of time. Therefore, a first and natural candidate can be the Hubble parameter $H$ multiplied with the energy density. Following Amendola et al. [56] and Guo et al. [57], we consider

$$Q = 3H\sigma\rho^m, \tag{40}$$

where $\sigma > 0$ is coupling coefficient which can be considered as a constant or function of redshift $z$.

Using equation (38), we obtain the energy density of dark matter as

$$\rho^m = \rho_0^m a^{-3(1+\omega^m-\sigma)} \tag{41}$$
$$= \rho_0^m (1+z)^{3(1+\omega^m-\sigma)},$$

where $\rho_0^m$ is an integrating constant.

Using equation (41) in equation (26), we obtain the energy density of DE as

$$\rho^{de} = DH^2\alpha\, a^\beta - \rho_0^m a^{-3(1+\omega^m-\sigma)} - K\alpha^{-1} a^{-(\beta+6)} \tag{42}$$
$$= \left(1+(1+z)^2\right)\left(D\alpha (1+z)^{-\beta} - 3\Omega_0^m (1+z)^{3(1+\omega^m-\sigma)}\right) - K\alpha^{-1}(1+z)^{(\beta+6)}.$$

Using equations (41) and (42) in equation (8), we obtain

$$\omega^{de} = \frac{H^2\left[(\beta+2)q - \left(\beta^2\left(1+\frac{\overline{\omega}}{2}\right)+\beta+1\right)\right]\alpha\, a^\beta + K\alpha^{-1} a^{-(\beta+6)} - 3H^2\omega^m\Omega_0^m a^{-3(1+\omega^m-\sigma)}}{H^2\left(D\alpha\, a^\beta - 3\Omega_0^m a^{-3(1+\omega^m-\sigma)}\right) - K\alpha^{-1} a^{-(\beta+6)}}. \tag{43}$$

This is the general form of the EoS parameter of DE in BD theory for interacting case. Here $q = -\frac{\ddot{a}}{aH^2}$ is the deceleration parameter.

Now using equation (13) in equation (43), we obtain the EoS parameter in terms of redshift as

$$\omega^{de} = -\frac{\left[(\beta+2)\dfrac{1}{1+(1+z)^2} + \left(\beta^2\left(1+\dfrac{\bar{\omega}}{2}\right)+\beta+1\right)\right]\alpha(1+z)^{-\beta} - K\alpha^{-1}\dfrac{(1+z)^{(\beta+6)}}{1+(1+z)^2} + 3\omega^m \Omega_0^m (1+z)^{3(1+\omega^m-\sigma)}}{D\alpha(1+z)^{-\beta} - K\alpha^{-1}\dfrac{(1+z)^{(\beta+6)}}{1+(1+z)^2} - 3\Omega_0^m(1+z)^{3(1+\omega^m-\sigma)}}.$$

(44)

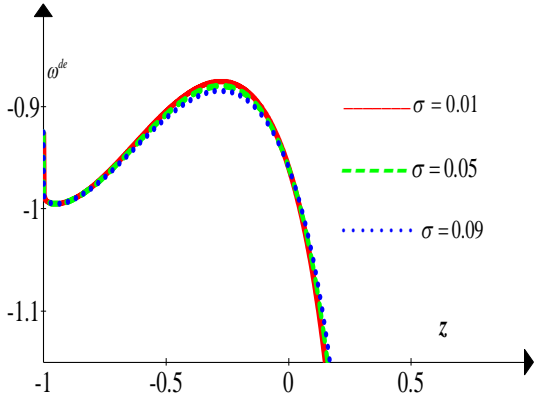
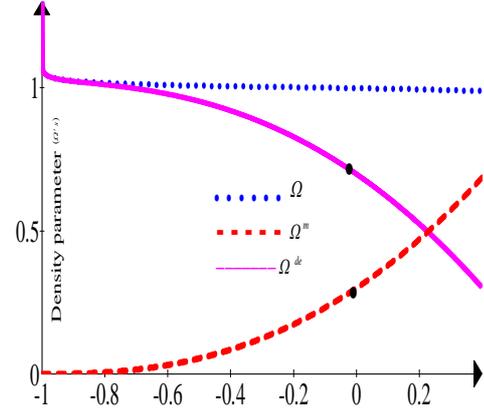

Fig.3: The plot of EoS parameter $\omega^{de}$ versus redshift $z$ for $\alpha = 1, \beta = 0.01, \Omega_0^m = 0.3$, $\omega^m = 0, \bar{\omega} = 2$ and vary $K$=0.01, 0.05, 0.09 resp.

Fig.4: The plot of total energy densities versus redshift $z$ for $\alpha = 1, \beta = 0.01$, $\Omega_0^m = 0.3$, $\omega^m = 0$ $\bar{\omega} = 2$, $K = 0.01$ and $\sigma = 0.18$.

The behavior of EoS parameter in terms of redshift $z$ is depicted in Fig. 3. We fixed the parameters $\omega^m = 0, K = 0.01$ and vary $\sigma$ as 0.01, 0.05 and 0.09 respectively. This figure shows that for all values of coupling constant $\sigma$, the EoS parameter of DE is varying in quintessence region, crosses the PDL and varies in phantom region. At late time (i.e. at $z = -1$), the EoS parameter of DE $\omega^{de} \approx -1$. Therefore we say that the cosmological constant is a suitable candidate to represent the behavior of DE in the derived model at late times.

The expression for matter energy density $\Omega^m$ and dark energy density $\Omega^{de}$ are given by

$$\Omega^m = \frac{\rho^m}{3H^2}$$
$$= \Omega_0^m(1+z)^{3(1+\omega^m-\sigma)},$$

(45)

and

$$\Omega^{de} = \frac{\rho^{de}}{3H^2}$$

$$= \frac{D\alpha}{3}(1+z)^{-\beta} - \frac{K\alpha^{-1}(1+z)^{(\beta+6)}}{3(1+(1+z)^2)} - \Omega_0^m(1+z)^{3(1+\omega^m-\sigma)}. \tag{46}$$

Using equations (45) and (46), we obtain total energy density parameter

$$\Omega = \Omega^m + \Omega^{de} = \frac{D\alpha}{3}(1+z)^{-\beta} - \frac{K\alpha^{-1}(1+z)^{(\beta+6)}}{3(1+(1+z)^2)}, \tag{47}$$

this equation is same as equation (37). The variation of density parameter $\Omega^m$ and $\Omega^{de}$ with redshift $z$ is depicted in Fig. 4. In figure the dot denotes the current value of these parameters. Hence we observed that in interacting case the density parameter has the same properties as in non-interacting case. From Figures 2 and 4, we observed that with interaction DE and DM follow one another. This means that in the recent history of the universe DE is being transformed into DM and the fluctuations do get more effective in the past. Note that the stronger the interaction, the more effectively structures will have been formed in the past.

## 4. Physical acceptability and stability analysis

To find the stability condition of corresponding models, we use squared speed of sound ($v_s^2$). A positive value of squared speed of sound ($v_s^2$) represents a stable model whereas the negative value of squared speed of sound ($v_s^2$) indicates the instability of model. A squared speed of sound ($v_s^2$) is defined as

$$v_s^2 = \frac{\dot{p}^{de}}{\dot{\rho}^{de}}. \tag{48}$$

The squared speed of sound for non-interacting and interacting models are respectively given by

$$v_s^2 = \frac{\left\{\begin{array}{l}\alpha\beta(z+1)^{-1-\beta}\left(1+(1+z)^2\right)\left[\beta^2\left(\frac{\bar{\omega}}{2}+1\right)+\frac{\beta+2}{1+(1+z)^2}+\beta+1\right]-\alpha(2z+2)(1+z)^{-\beta}\left[\beta^2\left(\frac{\bar{\omega}}{2}+1\right)+\frac{\beta+2}{1+(1+z)^2}+\beta+1\right]\\ +\frac{\alpha(\beta+2)(2z+2)(1+z)^{-\beta}}{1+(1+z)^2}+K\alpha^{-1}(\beta+6)(1+z)^{(\beta+5)}-3\omega^m\Omega_0^m(1+z)^{3\omega^m+2}\left[3\omega^m\left(1+(1+z)^2\right)+5z^2+10z+8\right]\end{array}\right\}}{\left\{\begin{array}{l}-\alpha D(z+1)^{-1-\beta}\left[2(1+z)^2-\beta\left(1+(1+z)^2\right)\right]-K\alpha^{-1}(\beta+6)(1+z)^{(\beta+5)}\\ -3\Omega_0^m(1+z)^{3\omega^m+2}\left[3\omega^m\left(1+(1+z)^2\right)+5z^2+10z+8\right]\end{array}\right\}} \tag{49}$$

and

$$v_s^2 = \frac{\left\{\begin{array}{l}\alpha\beta(z+1)^{-1-\beta}\left[1+(1+z)^2\right]\left[\beta^2\left(\dfrac{\overline{\omega}}{2}+1\right)+\dfrac{\beta+2}{1+(1+z)^2}+\beta+1\right]-\alpha(2z+2)(1+z)^{-\beta}\left[\beta^2\left(\dfrac{\overline{\omega}}{2}+1\right)+\dfrac{\beta+2}{1+(1+z)^2}+\beta+1\right]\\ +\dfrac{\alpha(\beta+2)(2z+2)(1+z)^{-\beta}}{1+(1+z)^2}+K\alpha^{-1}(\beta+6)(1+z)^{(\beta+5)}-3\omega^m\Omega_0^m\left[(3\omega^m-3\sigma+3)\left(1+(1+z)^2\right)(1+z)^{3\omega^m-3\sigma+2}+(2z+2)(z+1)^{3\omega^m-3\sigma+3}\right]\end{array}\right\}}{\left\{\begin{array}{l}-\alpha D(z+1)^{-1-\beta}\left[2(1+z)^2-\beta\left(1+(1+z)^2\right)\right]-K\alpha^{-1}(\beta+6)(1+z)^{(\beta+5)}\\ -3\Omega_0^m\left[(3\omega^m-3\sigma+3)\left(1+(1+z)^2\right)(1+z)^{3\omega^m-3\sigma+2}+(2z+2)(z+1)^{3\omega^m-3\sigma+3}\right]\end{array}\right\}}$$

(50)

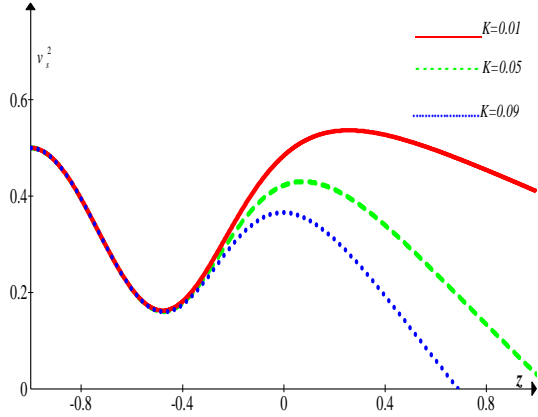

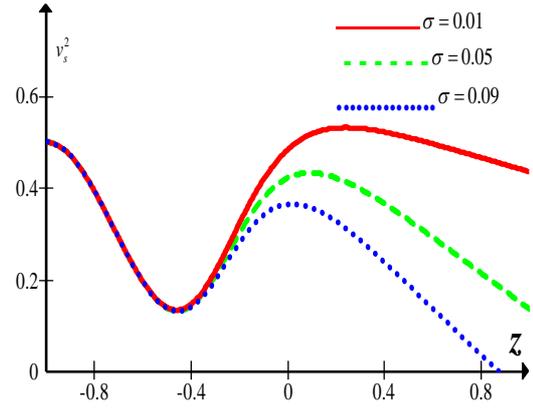

Fig.5: The plot of sound speed $v_s^2$ versus redshift $z$ in non-interacting case for $\alpha=1, \beta=1, \Omega_0^m=0.3, \overline{\omega}=1, \omega^m=0$ and vary $K$=0.01, 0.05, 0.09 resp.

Fig.6: The plot of sound speed $v_s^2$ versus redshift $z$ in interacting case for $\alpha=1, \beta=1$, $\Omega_0^m=0.3, \overline{\omega}=1, \omega^m=0, K=0.01$ and vary $\sigma$=0.01, 0.05, 0.09 resp.

From Figs. 5 and 6, it is observed that in our non-interacting and interacting models the sound speed remains positive (i.e. $v_s^2>0$), hence our models shows the stability throughout the evolution of the universe.

Secondly, the plot of weak energy condition (WEC), dominant energy condition (DEC) and strong energy condition (SEC) for non-interacting and interacting cases as shown in Fig. 7 and Fig. 8 respectively. It is observed that in non-interacting and interacting cases, the energy conditions obey the following restrictions:

(i) $\rho^{de}\geq 0$,

(ii) $\rho^{de}+p^{de}\geq 0$,

(iii) $\rho^{de}+3p^{de}\leq 0$.

From Figs. 7 and 8 and above expressions, we observed that the WEC and DEC are satisfied for non-interacting and interacting cases whereas the SEC is violated in entire evolution of the universe in non-interacting and interacting scenario.

Therefore, on the basis of above discussion and analysis, our corresponding models are physically acceptable.

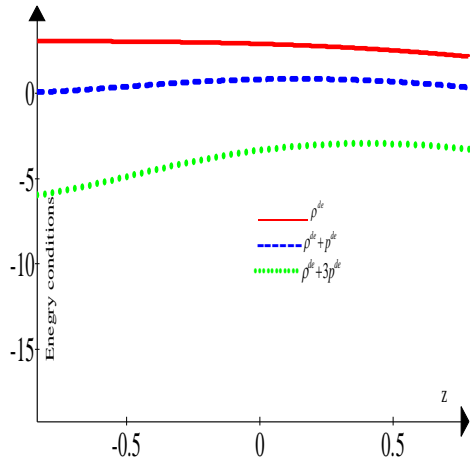

Fig.7: The plot of the weak $\rho^{de} \geq 0$, dominant $\rho^{de} + p^{de} \geq 0$ and strong $\rho^{de} + 3p^{de} \geq 0$ energy conditions versus redshift $z$ for non-interacting scenario.

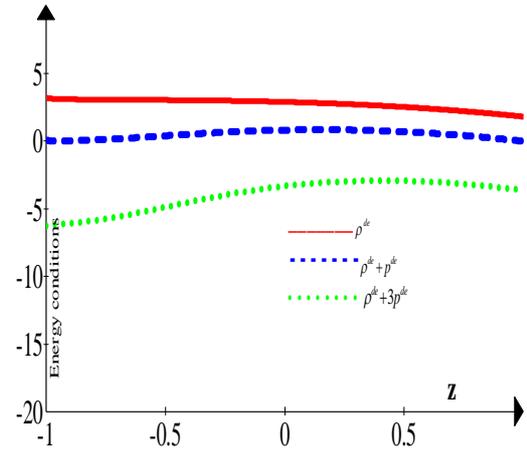

Fig.8: The plot of the weak $\rho^{de} \geq 0$, dominant $\rho^{de} + p^{de} \geq 0$ and strong $\rho^{de} + 3p^{de} \geq 0$ energy conditions versus redshift $z$ for interacting scenario.

## 5. Anisotropy parameter $(\Delta)$

The anisotropy parameter of expansion $(\Delta)$ is defined as

$$\Delta = \frac{1}{3H^2} \sum_{i=1}^{3} (H_i - H)^2$$

$$= \frac{(b_1^2 + b_2^2 + b_3^2)}{3H^2 \phi^2 a^6} = \frac{(b_1^2 + b_2^2 + b_3^2)(1+z)^{2(\beta+3)}}{3\alpha^2 (1+(1+z)^2)}.$$

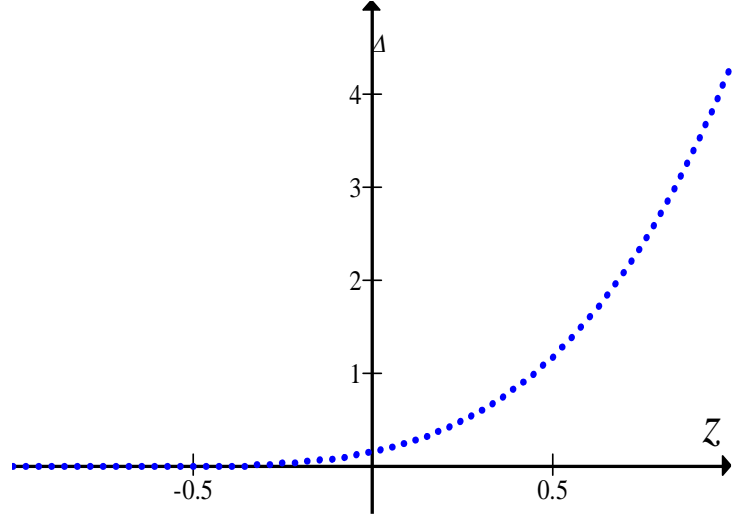

Fig.9: The plot of anisotropy parameter $(\Delta)$ versus redshift $z$

The anisotropy parameter versus redshift $z$ is depicted in fig.9. It is observed that the anisotropy parameter is decreasing function and dies out at late time $(z=-1)$. Hence the model reaches to isotropy which matches with the recent observations as the universe is isotropic at large scale.

## 6. Statefinder parameters

In order to get an accurate analysis to discriminate among the dark energy models Sahni et al. [58] proposed a new geometrical diagnostic named as statefinder pair $\{r,s\}$ which is constructed from scale factor (*a*) and its derivative upto third order. The statefinder parameters are defined as

$$r = \frac{\dddot{a}}{aH^3} \text{ and } s = \frac{r-1}{3\left(q-\frac{1}{2}\right)}. \tag{51}$$

These parameters allow us to characterize the properties of dark energy. Using these parameters one can describe the well-known region as follows: $(r,s)=(1,0)$ indicates $\Lambda CDM$ limit and $(r,s)=(1,1)$ indicates CDM limit, while $s>0$ and $r<1$ corresponds to region of phantom and quintessence dark energy era. The relation between statefinder parameters for our models are obtain as

$$r = \frac{1}{1+(1+z)^2}, \tag{52}$$

$$s = \frac{2(1+z)^2}{3\left(1+(1+z)^2\right)}. \tag{53}$$

From equations (52) and (53) a relation between parameters *r* and *s* is given by

$$s = \frac{2(1-r)}{3}. \tag{54}$$

The variation of parameter *s* versus *r* is plotted in Fig. 10.

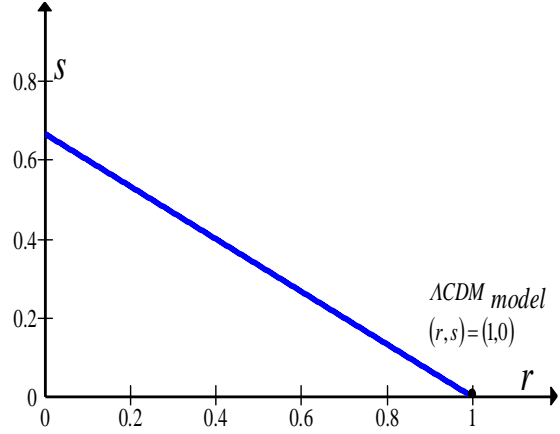

Fig.10: The plot of statefinder parameters *s* versus *r*.

From Fig. 9, it is seen that the curve passes through the point $(r=1, s=0)$, thus it can be concluded that our model corresponds to $\Lambda CDM$ model.

## 7. Conclusion

In this paper, we have studied interacting and non-interacting DE and DM in the anisotropic Bianchi type-I universe in the framework of Brans-Dicke theory of gravitation. To obtain the exact solutions of Brans-Dicke field equations we have used (i) the power-law relation between '$\phi$' and '$a$' and (ii) the average scale factor in terms of redshift. In non-interacting and interacting cases the general form of EoS parameter is derived. Then using the scale factor in terms of redshift, results are examined. We have discussed the physical acceptability and stability of our models. It is found that our models are physical acceptable and stable.

In non-interacting case, the EoS parameter of DE varying in quintessence region and crossing PDL depending on the values of constant *K* whereas in interacting case for all small values of coupling constant $\sigma$ the EoS parameter of DE varies in quintessence region, it crosses the PDL and varies in phantom region. However, at late time (i.e. at $z=-1$) the EoS parameter of both the cases tends to a cosmological constant $(\omega^{de} = -1)$. The anisotropy parameter of expansion $(A)$ is calculated and studied. It is observed that for $z=1$ (i.e. at initial time) it is infinite but at $z=0$ (i.e. at present time) it is closer to zero and at $z=-1$ it completely dies out which matches with the present day observations. Which indicate that our model at present time $(z=0)$ is closer to the de Sitter model and attains complete de Setter model at $z=-1$. i.e. we can say that the Bianchi type-I space-time reduces to flat FRW (isotropic) soon after the inflation. As mentioned in Carroll et al. [59], this ensures that there is no Big rip singularity; rather, the universe eventually settles into a de Sitter phase. Finally the statefinder diagnostic pair {$r, s$} is adapted to differentiate the different forms of DE. The trajectories in the {$r, s$} plane corresponds to the $\Lambda CDM$ model (as shown in Figure 10).